\documentstyle[twocolumn,prl,aps,epsfig]{revtex}

\newcommand{\ga}{\gamma}

\newcommand{\om}{\omega}

\newcommand{\beq}{\begin{equation}}
\newcommand{\eeq}{\end{equation}}
\newcommand{\ba}{\begin{array}}
\newcommand{\bea}{\begin{eqnarray}}
\newcommand{\ea}{\end{array}}
\newcommand{\eea}{\end{eqnarray}}

\newcommand{\ovl}{\overline}

\newcommand\comment[1]{ \hbox{[{\it Comment suppressed here.}\/]} }
\newcommand\hide[1]{}

\newcommand{\skipover}[1]{}

\begin{document}                                                
\title{Gapless Color Superconductivity}
\author{Mark Alford, J\"urgen Berges and Krishna Rajagopal}
\bigskip
\address{
Center for Theoretical Physics\\
Massachusetts Institute of Technology, Cambridge, MA 02139, USA
}
\date{MIT-CTP-2889}
\maketitle
\begin{abstract} 
%************
We present the dispersion relations for
quasiparticle excitations about the
color-flavor locked ground state of QCD 
at high baryon density.
In the presence of condensates which pair light and strange
quarks there need not be an energy gap in 
the quasiparticle spectrum.
This raises the possibility of
gapless color superconductivity, with
a Meissner effect but no minimum excitation energy. 
Analysis within a toy model suggests
that gapless color superconductivity
may occur only as a metastable phase.
%************
\end{abstract}
\pacs{}
\begin{narrowtext}
%\section{Introduction}
Strongly interacting matter at sufficiently high baryon density
and low temperature is in a color superconducting state
characterized by a condensate of quark Cooper 
pairs \cite{Barrois,BailinLove,ARW2,RappETC}.
Such a condensate gives mass to
gauge bosons via the Anderson-Higgs mechanism.  
In addition to the Meissner effect, 
a superconducting phase is typically characterized
by an energy gap $2\Delta$  
in the density of quasiparticle states. The gap
corresponds to the minimum energy necessary to excite one
quasiparticle pair relative to the ground
state energy.  In a typical superconductor,
the Meissner effect is therefore accompanied by 
the characteristic thermodynamic
consequences of a gap, like a specific heat 
$C_V\sim e^{-\Delta/T}$.
In this letter we argue that in a color superconductor which
includes pairing between quarks with differing mass, an
energy gap is not mandatory despite the presence of a
condensate. Thus, one may have {\it gapless superconductivity} 
in QCD.

At sufficiently high density, quark
matter is in the {\it color-flavor locked} state (CFL) which
involves pairing of the light up and down quarks and the 
heavier strange quark ($u$, $d$ and $s$) \cite{CFL,SW1,ABR,SW2}.
In this introduction, we give 
a model independent argument that for pairing of light with
strange quarks
the energy gap in the quasiparticle spectrum vanishes if the
condensate $\langle us \rangle$ (or 
$\langle ds \rangle$) is less than of order 
$M_s^2/4\mu$. Here, $\mu$ is the quark number chemical potential, 
$M_s$ is the $\mu$-dependent effective, or constituent,
strange quark mass and $\langle us \rangle$ denotes the proper 
self energy. This raises the possibility of
gapless color superconductivity if the condensate 
$\langle us \rangle$ is nonzero
and sufficiently small in high density QCD.

We begin with the dispersion relations for two noninteracting fermions,
one massless $(u)$ and the other $(s)$ with mass $M_s$ \cite{masslessu}.  
At nonzero $\mu$,
the $u$ dispersion relation is $\omega_u=\pm(|{\bf q}|\pm \mu)$
and the $s$ dispersion
relation is $\omega_s=\pm(\sqrt{|{\bf q}|^2+M_s^2}\pm \mu)$, 
where we are measuring energy relative
to the Fermi energy. In Fig.~1 we plot the positive branches,
corresponding to empty states which can be filled by 
single particle excitations. 
The two different Fermi momenta are apparent, as are
the dispersion relations for $u$ and $s$ quarks at momenta
greater than their respective Fermi momenta and for 
$u$ and $s$ holes at momenta less than their respective Fermi momenta.
The beginning of the $u$ anti-particle branch is also visible.
\begin{figure}
\begin{center}
\epsfig{file=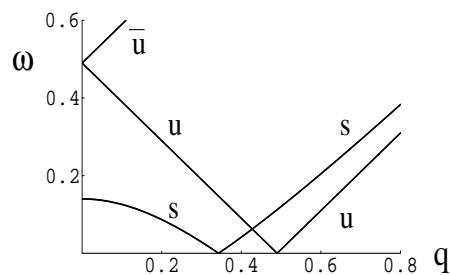,width=2.3in,height=1.40in}
\end{center} \label{fig1}
\caption{Dispersion relations for free massless $u$
quarks and $s$ quarks with mass $M_s=350$ MeV at $\mu=490$ MeV.
}
\end{figure}
The $s$ and $u$ dispersion relations cross at $\omega_u = \omega_s =
M_s^2/4\mu$.  We expect this degeneracy to be lifted in the
presence of interactions between $u$ and $s$ quarks.  Just as 
a $\langle uu \rangle$ condensate would open a gap $2\langle uu \rangle$ 
at the $u$ Fermi surface, we expect a $\langle us \rangle$
condensate to open a ``gap'' as shown in Fig.~2.
\begin{figure}
\begin{center}
\epsfig{file=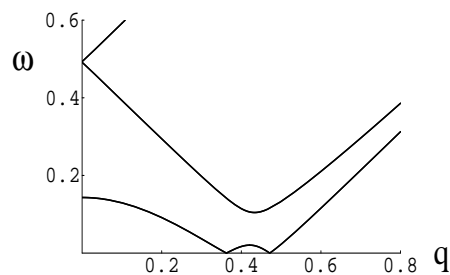,width=2.3in,height=1.40in}
\end{center} \label{fig2}
\caption{Dispersion relations for massless $u$ quarks and strange
quarks with $M_s=350$ MeV at $\mu=490$ MeV in the presence
of a $\langle us \rangle = 45$ MeV condensate.} 
\end{figure}
The ``$u$-hole branch'' and ``$s$-particle branch'' are 
separated by a ``gap'' of $2\langle us \rangle$, 
but this ``gap'' is {\it not} at the Fermi energy ($\om=0$).
This figure depicts a gapless
superconductor: $\langle us\rangle\neq 0$, but there are
clearly quasiparticle excitations with $\omega=0$. 
In Fig.~2, $\langle us\rangle < M_s^2/4\mu$.  
For larger
values of the condensate, the picture
must change. We will see that for
$\langle us\rangle \gtrsim M_s^2/4\mu$, there is in fact a true gap,
as shown in Fig.~3.  
\begin{figure}
\begin{center}
\epsfig{file=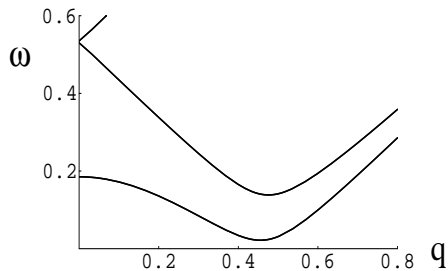,width=2.3in,height=1.40in}
\end{center}
\caption{As in Fig.\ 2, with $\mu=525\!$ MeV and 
$\langle us\rangle=85\!$ MeV.}
\end{figure} 
For still larger values of the condensate,
we find the dispersion relations of Fig.~4.  
Note that in these cases, although the superconductor is not gapless,
the two dispersion relations have quite different minima, with
one gap  smaller and the other one larger than
the value of the condensate itself.
\begin{figure}
\begin{center}
\epsfig{file=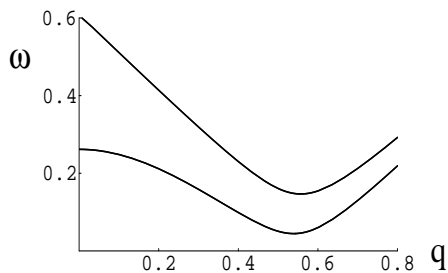,width=2.3in,height=1.40in}
\end{center}
\caption{As in Fig.\ 2, 
with $\mu\!=\!600\!$ MeV and $\langle us\rangle\!=\!100\!$ MeV.}
\end{figure}

The qualitative lessons of Figs.~1-4 are generic but the 
figures themselves were derived in a particular model. In
Section II, we give a quantitative explanation
of the relationship between condensates and 
the dispersion relations and gaps illustrated in the Figures.
In Section III, we present the model within which the specific
parameter values used in the Figures are derived.
The remaining question, then,
is whether a phase of QCD exists in which $\langle us \rangle$
is nonzero but sufficiently small that gapless
superconductivity arises as in Fig.~2.  We have argued
in \cite{ABR} that if $\langle us\rangle$ 
is less than the mismatch in Fermi momenta $p_F^u-p_F^s\sim
M_s^2/2\mu$ then the $\langle us\rangle$ condensate vanishes
at a first order unlocking transition.
Because $M_s^2/4\mu < M_s^2/2\mu$, these qualitative arguments
suggest that a gapless superconducting phase is never the
thermodynamic ground state.  
This is consistent with what we find in Section III:
in our model, gapless superconductivity only arises as a metastable
state. However, neither the model nor the qualitative arguments
should be trusted to within a factor of two,
and it remains to be seen whether QCD admits a stable gapless
superconductor CFL phase near the unlocking transition.
We close in Section IV with 
implications for the physics of the CFL
phase, and lessons which can be learned from analogue systems
in which gapless superconductivity may arise.

\setcounter{section}{1}
\section{From Condensates to Dispersion Relations and Gaps}
It is convenient to write the 
free Euclidean inverse quark propagator $G_0^{-1}$
at nonzero 
$\mu$ as a matrix acting on the 
column vector ${\psi \choose \bar{\psi}^T}$ 
\bea
G_0^{-1} = \left(\begin{array}{cc}
& C (q_{\nu}\ga^{\nu}+ i\mu \gamma^4) C \\ 
q_{\nu}\ga^{\nu}- i\mu \gamma^4 & 
\end{array}\right)
\eea 
where $C$ is the charge conjugation matrix.
We write the full propagator of the
interacting fermion system 
as 
\begin{equation}
G = (G_0^{-1} + \Sigma)^{-1} \, ,
\label{fullprop}
\end{equation}
with the proper self energy
\bea
\Sigma = \left(\begin{array}{cc}
 \langle \psi \psi \rangle & 
-i\, \langle \overline{\psi} \psi \rangle \\ 
i\, \langle \overline{\psi} \psi \rangle & 
\langle \psi \psi \rangle
\end{array}\right)\, . \label{sig}
\eea

As a simple example, to demonstrate the physics
in Figs.~1-4, consider
two species of fermions forming the Lorentz scalar condensates
\begin{equation}
\langle \psi \psi \rangle = C \gamma_5 \left(\begin{array}{cc}
0 & f \\ 
f & 0
\end{array}\right)\, \, , \quad 
\langle \overline{\psi} \psi \rangle = \left(\begin{array}{cc}
0 & 0 \\ 
0 & M_s 
\end{array}\right) \, .
\label{ansatz}
\end{equation}
The two fermions should be thought of as a massless $u$-quark 
with a certain color and 
an $s$-quark of a different color with constituent mass $M_s$.
We will see in the next section that the ansatz (\ref{ansatz})
suffices to describe those features of the CFL phase of interest to us.
The condensate $f$ can be obtained by solving a self-consistent
Schwinger-Dyson equation for $\Sigma$. Assuming
a four-fermion interaction, this is given schematically by
\begin{center}
$\displaystyle{{ \Sigma} \quad = \quad }$
\parbox{1.in}{
\epsfig{file=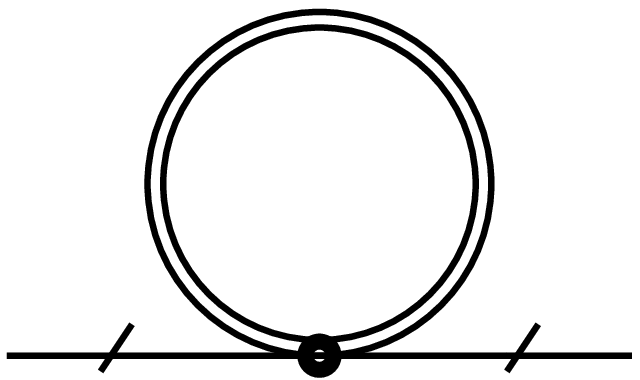,width=0.6in}
}
\end{center}
where the loop denotes a momentum integration
over the full propagator (\ref{fullprop}) and where
the external legs have been amputated.

In order to determine the gap associated with a given 
condensate $f$, we 
need the quasiparticle dispersion 
relations \cite{plasmino}. These are determined by
the poles of the full propagator (\ref{fullprop}). After
some algebra one finds that
the dispersion relations $\omega({\bf q})$
are given by solutions of
$D_+ =0$  or $D_- =0$, upon noting that $\omega=iq_0$, 
with
\bea
D_{\pm}\, &=& \, \left[ q_0^2+ (\mu - |{\bf q}|)^2 + f^2) \right]
\left[ q_0^2+ (\mu + |{\bf q}|)^2 + f^2) \right] \nonumber\\ 
&& \displaystyle{- M_s^2 \left[ i q_0 \pm (\mu - |{\bf q}|) \right]
\left[ i q_0 \pm (\mu + |{\bf q}|) \right]} \, . \label{den}
\eea

It is instructive to consider the massless case,
$M_s \to 0$. The dispersion relation is then 
$\omega({\bf q}) = \pm \sqrt{(\mu \pm |{\bf q}|)^2 + f^2}$.
The first $\pm$ distinguishes filled negative energy states
from positive energy states describing excitations.
The second $\pm$ distinguishes the particle/hole branch 
from the antiparticle branch.  The particle/hole branch 
%$\omega({\bf q}) =  \sqrt{(\mu - |{\bf q}|)^2 + f^2}$
has a minimum of $\omega(\mu)=f$. There is a gap $2\Delta = 2f$
at the Fermi surface. We thus recover the familiar result
that the gap and the condensate are equal if we pair quarks
whose masses are degenerate.  For $M_s\neq 0$, we obtain
dispersion relations as illustrated in Figs.~1-4.  
We thus observe that when quarks with different masses
pair, the condensate $f$ yields quasiparticles
with two different dispersion relations. The gaps for the two
branches
differ from each other and from $f$. 
For small 
enough $f$, gapless superconductivity results.

In the presence of any diquark condensate, the quasiparticles are
linear combinations of particles and holes.  In the presence
of a $\langle us \rangle$ condensate, they are also linear
combinations of $u$ and $s$. We illustrate this by 
diagonalizing the propagator matrix (\ref{fullprop}),
computing its eigenvectors, and thus determining the 
probability that the quasiparticles in Fig.~3 are $u$ (particles
or holes).  The results, plotted in Fig.~5, demonstrate
that the upper (lower) branch is mostly $u$ ($s$) holes
at low $|\bf{q}|$ and mostly $s$ ($u$) particles at high $|\bf{q}|$,
as must be the case given the sequence of Figs.~1-4.
This allows us to explain the transition from Fig.~2 to Fig.~3
more clearly.  In Fig.~2, because 
$\langle uu \rangle=\langle ss \rangle=0$ there is no gap at the 
Fermi surface.  In Fig.~3, the positive and negative energy branches
which are close to touching at the Fermi surface both 
describe linear combinations of $u$ and $s$. This means
that the $\langle us \rangle$ condensate keeps them apart,
and a true gap opens up.
\begin{figure}[h]
\begin{center}
\epsfig{file=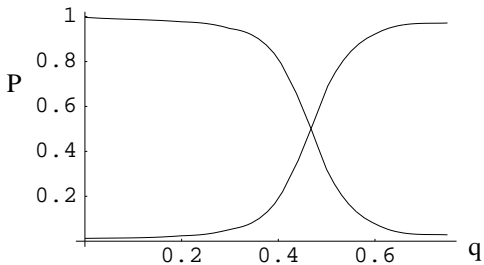,width=2.5in} %height=1.3in
\end{center}
\caption{Probability that a quasiparticle is $u$ as opposed to $s$,
for the lower two dispersion relations in Fig.~3.}
\end{figure}

\section{Metastable Gapless Superconductivity}

We follow Ref.\ \cite{ABR} and describe superconducting 
strange quark matter in a model in which quarks interact 
via a four-fermion interaction abstracted from single-gluon 
exchange,
${\cal L}_{\rm int} = \, G\int d^4 x \left( \ovl \psi \lambda^a 
\gamma^\mu \psi\right)\left( \ovl \psi \lambda^a 
\gamma_\mu \psi\right)$,
with $G$ chosen to give reasonable vacuum physics \cite{ABR}.
At asymptotically high densities, such phenomenological approaches 
have been superseded by calculations done using QCD itself \cite{Son}.  
As conjectured in \cite{ABR}, recent work\cite{SW3,PR}
demonstrates that at accessible densities the two approaches 
are in reasonable agreement in
their predictions for the magnitude of the condensates.

At sufficiently high density, QCD is in the CFL phase 
which, for $M_s\neq 0$, is characterized by  an unbroken
$SU(2)_{{\rm color}+L+R}$ symmetry describing simultaneous
$SU(2)$ rotations in color and vector flavor.
In this phase,
all nine quarks (3 colors times 3 flavors)
form condensates.  Thus, the ansatz (\ref{ansatz}) should
be replaced by the $9\times 9$ block-diagonal matrix of  Ref.\ \cite{ABR}. 
Four of the
nine quarks form two doublets under $SU(2)_{{\rm color}+L+R}$
and pairing between elements of these doublets results in 
$\langle us\rangle$ and $\langle ds\rangle$ condensates
described by two $2\times 2$ blocks, 
each with the form (\ref{ansatz}).  
Diagonal $\langle uu \rangle$ and $\langle ss \rangle$
entries in (\ref{ansatz}) do not arise because they 
break the $SU(2)_{{\rm color}+L+R}$ symmetry; 
their presence would preclude gapless superconductivity.
Three of the remaining five quarks, linear combinations
of $u$ and $d$ only, pair among themselves 
and the resulting
dispersion relations have gaps equal to the
associated $\langle ud \rangle$ condensate.
In the CFL phase,
the last $2\times 2$ block does involve $\langle us \rangle$
pairing but its entries 
are such that no dispersion relation can become gapless \cite{ABR}.
%The last two quarks are each linear combinations of $u$, $d$, and $s$,
%and are each singlets under $SU(2)_{{\rm color}+L+R}$ 
%which pair with themselves. For these quarks, condensates in all 
%channels are present, including $\langle uu \rangle$, 
%$\langle dd \rangle$ and $\langle ss \rangle$, consistent 
%with the $SU(2)_{{\rm color}+L+R}$ symmetry.
The only blocks from the full $9 \times 9$ ansatz of Ref.\ \cite{ABR}
which can yield gapless superconductivity are therefore the two copies
of the ansatz (\ref{ansatz}).

In Ref.\ \cite{ABR}, we have solved the coupled mean-field
Schwinger-Dyson equations for the CFL condensates
including $f$.  The only condensate of interest here
is $f$, and we plot it in Fig.~6. 
\begin{figure}[h]
\begin{center}
\epsfig{file=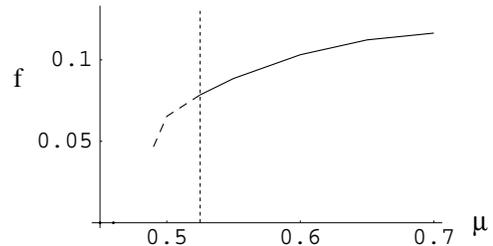,width=2.5in,height=1.3in}
\end{center}
\caption{$f$ in the CFL
phase as a function of $\mu$ for $M_s=350$ MeV. For $\mu<525$ MeV
the stable superconducting phase has $f=0$, and
the CFL phase exists only as a metastable phase 
as indicated by the dashed line.} 
\end{figure}
We can now explain the choice of parameters in Figs.~1-4.
Fig.~4 depicts the dispersion relations at a generic point
in the CFL phase, with $\langle us\rangle = f=100$ MeV 
taken from Fig.~5 at $\mu=600$ MeV.
Fig.~3 gives the dispersion relations {\it at} the unlocking
transition.  The gap ($15$~MeV) is rather small compared to the condensate 
($85$~MeV),
but is still nonzero.  At the unlocking transition, 
$\langle us \rangle=M_s^2/2.5 \mu$ \cite{ABR}. 
We find that gapless superconductivity
sets in for $\langle us \rangle \leq M_s^2/3.8 \mu$, in the metastable
phase. Fig.~2 shows the dispersion relations 
for the metastable CFL phase below
the unlocking transition at $\mu=490$ MeV, $\langle us \rangle = 45$ MeV.
In the model, as the simple arguments given in the introduction suggest,
gapless superconductivity does not occur as the thermodynamic
ground state. It does occur as a metastable phase.

\section{Implications and Analogues}

It is striking that the CFL phase breaks chiral symmetry \cite{CFL}
and has the same symmetries as
sufficiently dense hypernuclear matter\cite{SW1}, so there need be no phase
transition between them. It has been hypothesized \cite{SW1} that
there is ``quark-hadron continuity'': as $\mu$ is increased
the baryonic condensates change
continuously into quark condensates,
%
% eg, $\langle p\Xi^-\rangle$ and $\langle n \Xi^0\rangle$ become
% $\langle us \rangle$ \cite{ABR}, 
%
and the gaps at the hyperon Fermi surfaces become gaps at the quark
Fermi surfaces. This raises the possibility of obtaining information
about the hadronic phase from calculations in the weakly-coupled quark
phase\cite{SW1,ABR}. 
Our results indicate, however, that it is in practice
difficult to exploit 
quark-hadron
continuity in this way,
because the relative sizes
of the {\em gaps} in the various channels may be quite different in the
hadronic and quark regimes, even if the relative sizes of the 
{\em condensates} are similar.
The reason is that, as shown above, in channels
where there is pairing of fermions of different mass,
the gaps depend sensitively on the fermion masses
as well as on the condensates. However,
the fermion masses change dramatically in magnitude and in pattern
as $\mu$ is increased from the hypernuclear phase into the quark phase.
In the quark phase, there are six light quarks and three 
heavier strange quarks, which is quite unlike the 
pattern of masses for the baryons.
We conclude that it will be hard to infer physics of hadronic matter,
such as the ratio of the gap in one channel to that in another, 
from calculations performed in the quark matter regime. 

Pairing between fermions with different dispersion relations
occurs in other contexts.  Analogues include
pairing between neutrons and protons in nuclear matter which is
not isospin symmetric \cite{Sedrakian}, and pairing between spin-up
and spin-down electrons in an ordinary superconductor
placed in a magnetic field $H$ which introduces a Zeeman splitting
$\mu_B H$ \cite{ClCh,Tinkham}.
The latter case yields a particularly apt analogue.
In some superconductors in which
the spin-orbit coupling is small, gapless superconductivity
would set in for $\mu_B H > \Delta$, where $\Delta$ is the gap
at $H=0$ \cite{ClCh,Tinkham}. 
However, what happens instead is that a first order phase
transition to a nonsuperconducting phase occurs at 
$\mu_B H = \Delta/\sqrt{2}$ \cite{ClCh,Tinkham}.  
This is precisely analogous to
what we have found: unlocking precludes gapless superconductivity,
except as a metastable phase.  It is worth noting that 
if one 
introduces paramagnetic impurities, instead of a uniform
magnetic field, there is a range of impurity concentrations
for which the superconductor {\it is} gapless \cite{AbGor,Tinkham}.  
%This has been confirmed in tunnelling experiments.\cite{Tinkham}
This example suggests that although
in our model gapless superconductivity only occurs below the unlocking
transition, where the CFL phase is metastable, in QCD itself
it may occur for a range of chemical potentials above the unlocking
transition, where the CFL phase is the stable thermodynamic ground
state.

The thermodynamic
properties of quark matter in the CFL phase 
need not be characterized by a gap. 
Gapless superconductivity is most likely to occur at the lowest
densities at which the CFL phase is present. This makes
it of potential interest in neutron star phenomenology.
Further work is required to elucidate the importance of
this observation, because the quasiparticles which we have
analyzed are not the only excitations in the CFL
phase.  The Nambu-Goldstone bosons arising from the spontaneous
breaking of global symmetries (chiral symmetry and $U(1)_B$)
and the massive gluons arising from the breaking of gauge
symmetries will also contribute to the specific heat and
to transport properties. It remains to be seen how significant
the presence of quasiparticles with a gapless dispersion
relation would be in the context of neutron star phenomenology.

Research supported in part by the DOE under agreement DE-FC02-94ER40818.
The work of KR is supported in part by a DOE OJI grant and by the 
A. P. Sloan Foundation.

\end{narrowtext}


\begin{references}
\bibitem{masslessu}
Small $u$ and $d$ quark masses do not significantly affect our
results.
\end{references}
\end{document}